\documentclass{svmult} 

\usepackage{graphicx}        
\usepackage[bottom]{footmisc}
\usepackage{amssymb}

\newcommand{\lsim}   {\mathrel{\mathop{\kern 0pt \rlap
  {\raise.2ex\hbox{$<$}}}
  \lower.9ex\hbox{\kern-.190em $\sim$}}}
\newcommand{\gsim}   {\mathrel{\mathop{\kern 0pt \rlap
  {\raise.2ex\hbox{$>$}}}
  \lower.9ex\hbox{\kern-.190em $\sim$}}}

\begin{document}
\input epsf 
\title*{A Conceptual Tour About the  Standard Cosmological Model}
\author{Antonio L. Maroto \and Juan Ram\'{\i}rez}
\institute{Dept. F\'{\i}sica Te\'orica I, Universidad Complutense de 
Madrid, 28040 Madrid, Spain
}

\maketitle

\vspace*{-.4cm}

\section*{Acknowledgement}
This paper is dedicated to Prof. Alberto Galindo on occasion of his 70th
birthday. We are delighted to write on Cosmology, which is one of his
favourite subjects. He taught both of us at the Complutense University
in Madrid, and we would like to thank him for his outstanding teaching. 
 His insistence on logical reasoning, physical intuition 
and mathematical comprehension
of natural laws followed  the best tradition of Galilei: 
{\it the book of Nature is written in mathematical language}.

\abstract{With the beginning of the XXIst century, a physical model of our 
Universe, 
usually called the Standard Cosmological Model (SCM), is reaching an important
level of consolidation, based on accurate astrophysical data 
  and also on theoretical developments.
In this paper we review the interplay between the basic concepts and
observations underlying this model. The SCM is a complex and 
beautiful building, recieving inputs from many branches of physics.
Major topics reviewed are: General Relativity and the cosmological constant,
the Cosmological Principle and Friedmann-Robertson-Walker-Lema\^{\i}tre 
 models, Hubble diagrams and dark 
energy, large scale structure  and dark matter, 
the cosmic microwave background, Big Bang nucleosynthesis, and
inflation.}


\section{Introduction}
Cosmology is the science that studies the Universe as a whole. From the dawn 
of civilization mankind has asked questions about the structure and 
composition of the Universe and the laws that govern it. 
Examples of such questions are

\begin{itemize}
\item How old is the Universe?
\item What is the size and the geometry of the Universe? 
\item How did the Universe began and how will it end?
\item What is the composition of the Universe?
\item How did the matter and the structures that we observe in the Universe 
originate?
\end{itemize}

A very remarkable historical and philosophical feature of the present 
epoch, and in particular of the last few years, is the emergence of Cosmology
as a mature science in which most of the former questions have a precise
answer. For example, within a few percent, we know that the Universe is
13.7 Gyr. old, and that the part of it within our present horizon is flat
and has a radius of 14-15 Gpc \footnote{1pc = 3.2615 light-years = $3.0856 
\times 10^{13}$ km}. These definite answers to such fundamental questions 
are possible in the context of the Standard Cosmological Model (SCM),
that arised in the second half of the last century with the discovery of
the Cosmic Microwave Background (CMB). The SCM is presently growing 
towards a  level
of development and reliability comparable to the Elementary Particles
Standard Model (EPSM). This achievement has been possible due to 
the combined effort of theory  based on fundamental physics, and a large 
number of astronomical observations from a host of scientific satellite 
and balloon probes, and also from the ground. 

The SCM is based on five strong pillars: i) The General Theory of 
Relativity, introduced by A. Einstein in 1916, 
which provides the theory of the gravitational 
field and the basic framework for the cosmological models. 
ii) The Cosmological Principle, also introduced by Einstein in 1917, that
states the homogeneity and isotropy of the Universe. 
iii) The Hubble law, discovered
by E. Hubble in 1929,  stablishing the expansion of the Universe,  and
the Hubble diagrams which allow to determine the acceleration/deceleration
of the Universe by means of standard candles. iv) The CMB corresponding 
to blackbody radiation at $T = 2.725$ K, accidentally discovered by
A. Penzias and R. Wilson in 1964, whose mean isotropy
supports the cosmological principle, and whose small anisotropies in the
spatial distribution of the sky temperatures contain a wealth of information 
on the cosmological parameters. v) The light
elements cosmological abundances, namely $^1$H, $^2$D, $^3$He, $^4$He, 
and $^7$Li, originated during the primordial Big Bang Nucleosynthesis (BBN) 
when
the Universe was 100 s. old, whose theoretical analysis pioneered by G. Gamow 
in 1948, reinforces the Big Bang scenario. To this list it should be added the
analysis of the Large Scale Structure (LSS) in the Universe based on
galaxy catalogs, which has recently received a major boost with the 
Two degree Field Galaxy Redshift Survey (2dFGRS) and the 
Sloan Digital Sky Survey (SDSS). 

The elucidation of the impact of the former pillars in the model conveys
a lot of fundamental physics, from which the EPSM is not the lesser part.
From this elucidation emerges the SCM as the Hot Big Bang model 
with the addendum of a primordial inflationary phase. In the Hot Big Bang 
model, the Universe expands and cools from a very dense and hot state. The
temperature of the plasma in the early phase, and of the CMB photons later
on, scale as $T \propto a^{-1}$, with $T$ being the temperature and 
$a$ the scale factor of the Universe.
The history of this expansion and  cooling can be followed backwards in time
in terms of well known fundamental physics until a time $t \sim 10^{-5}$s
corresponding to a temperature $T \sim 10^{12}$K, when the Universe was
$10^{12}$ times smaller than it is now. This corresponds to a mean energy
in the plasma of 100 MeV, and although particle accelerators have explored
the EPSM up to energies of 1 TeV, the accurate control of the physics 
that is going on is lost at this point because of our present lack of knowledge
about the hadron to quark-gluon plasma phase transition, which ought to 
occur around 170 MeV. However the basic features of the model can 
be further extrapolated up to the Grand Unification temperature 
$T_{\mathrm{GU}} \sim  10^{29}$K ($t \sim 10^{-39}$s) 
and ultimately to the Planck temperature $T_{\mathrm{P}} \sim  10^{32}$K 
($t_{\mathrm{P}} \sim 10^{-44}$s)
where our lack of understanding of \emph{quantum gravity?}
prevents further 
extrapolation to earlier times. Classically, the extrapolation to zero
time would lead to a state of infinite temperature and density known 
as the Big Bang singularity. Thus, the name Big Bang has two meanings
in Cosmology. The first refers to the very hot and dense
plasma composed of protons, neutrons, electrons, positrons, neutrinos 
and photons existing at $t \sim 10^{-5}$s from which our present Universe,
containing ourselves, emerges through expansion, cooling and growing 
of structures. This Big Bang is firmly stablished through astronomical
observations related to CMB, BBN, LSS and Hubble diagrams. The second 
meaning refers to the classical singularity from which the Universe
seems to spring at Planck time $t_{\mathrm{P}} \sim 10^{-45}$s. A consistent
quantum mechanical description of this phase is unfortunately not 
yet available, but this is a fertile ground for promising theoretical
speculations like string theory, M-theory, branes and other TOEs (Theory
of Everything). 
 
In 1980 A. Guth and in 1981 A. Linde, inspired in Grand Unified Theories,
introduced  the hypothesis that around the epoch between Planck time 
and Grand Unification time, the Universe underwent a period of
rapid exponential expansion which augmented its size by a factor between
$e^{50}$ and $e^{70}$. This idea was introduced to solve the so-called
problems of the Hot Big Bang model, meaning to find a mechanism which
explains why the Universe is so smooth, old and flat. The introduction 
of the inflationary phase provides a compelling natural explanation,
but more importantly, it gives a model for structure formation based on
the quantum fluctuations of the inflaton field, whose ground state
energy drives the rapid exponential expansion of the Universe.    

In this short review we will present a broad view description of the 
Standard Cosmological Model, and of the theoretical ideas and observational 
facts that support it. An updated textbook on the Standard Cosmological
Model is for example \cite{Dodelson}.

\section{General Relativity, the Cosmological Principle and FRWL Models}

\subsection{General Relativity and the Cosmological Constant}

Despite its weakness with respect to other fundamental forces, the long 
range and the absence of screening make gravity the driving force of
the cosmos. To the extent of the present knowledge, the gravitational force is
correctly described by Einstein's General Relativity, which also
provides the geometrical framework for cosmological models. In the broad
view General Relativity consists of two basic elements: \emph{the Equivalence
Principle} and \emph{Einstein's field equations}. 

In its Newtonian version the Equivalence Principle amounts to the 
identification of the gravitational mass entering Newton's gravity law 
and the inert mass entering the second law of Newtonian mechanics.
The first precise experimental test of this equality was performed
by R. E\"{o}tv\"{o}s in 1890, who showed that the ratio 
$m_{\mathrm{g}}/m_{\mathrm{i}}$ does not differ from one substance to another 
in more than one part in $10^9$. The most accurate test that all bodies
fall with the same accelaration in a gravitational field comes
from the comparison of the accelerations of the Moon and the Earth as they
fall around the Sun by means of lunar laser ranging. These accelerations 
agree to an accuracy of $1.5 \times 10^{-13}$ \cite{Equiv}.  

Equality of the inertial and gravitational masses led Einstein  
to formulate the relativistic version of the equivalence principle 
by identifying the gravitational field with the metric tensor 
$g_{\mu \nu}$ describing the (pseudo) Riemannian geometry of the 
space-time manifold.

The second element of General Relativity are Einstein's field equations
for the gravitational field
\begin{equation}
R_{\mu\nu} - \frac{1}{2} g_{\mu\nu}R = 8\pi G T_{\mu \nu} + \Lambda g_{\mu \nu} ~~.
\label{Einstein}
\end{equation}
These equations derived by D. Hilbert and  A. Einstein in 1916, are similar 
in spirit to Maxwell's equation for the electromagnetic field, and relate
the geometry of space-time embodied in the Ricci tensor $R_{\mu\nu}$ 
and the scalar curvature $R$, with the stress-energy tensor 
$T_{\mu\nu}$ as its source.

The second term in the second member, containing the \emph{cosmological
constant} $\Lambda$,
was introduced by Einstein only in 1917, to  
obtain his famous static cosmological solution. After the discovery of 
the expansion of the Universe by E. Hubble in 1929, Einstein considered 
the introduction of the  $\Lambda$ term as the biggest mistake of his life.
This case provides a good example of the (dangerous) role that 
preconceptions can play in Cosmology. The idea of a stationary Universe 
was the current paradigm, which had reigned for centuries,
when Einstein introduced General Relativity. The idea of \emph{evolution},
a common place for Life Sciences since the nineteenth century, 
was taken seriously in Cosmology only after the discovery of the CMB in 1964.

Ironically enough, after long theoretical efforts to prove that $\Lambda$
should be equal to zero, the $\Lambda$ term reappears in two places in 
modern Cosmology: during the inflationary phase as the 
vacuum energy of the inflaton field, and as a natural explanation
for the observed acceleration of the Universe, discovered in 1998 
\cite{SN}. In the modern view of Quantum Field Theory, the cosmological
constant should receive a contribution from the vacuum energy or zero
point energy of the oscillatory modes of the quantum fields. Then, in terms 
of the energy density $\Lambda/8\pi G$, a natural value would be 
$M_p^4 \sim 10^{76}$ GeV$^4$. This value is even unacceptable for
the inflationary phase of the Universe, and it is disparately wrong by 
124 orders of magnitude when compared with a cosmological constant driving the
present acceleration of the Universe. In addition, each phase transition,
like the GUT or the electroweak, shoud give a jump in the vacuum energy 
density of the order of $E^4$, with $E$ being the energy scale of the 
transition. This rather odd state of affairs with $\Lambda$, 
is undoubtely one of the most important problems that
needs to be settled in physics. 

On the other hand, the present value of the cosmological constant, 
although very 
important for cosmological dynamics, is too small to have any effect
at non-cosmic distances. Indeed 
Einstein's equations without $\Lambda$ have been experimentally checked
in the weak fields of the Sun and the Earth. In this regime, the simplest
Einstein's equations without $\Lambda$ provide a correct description of
gravity. However, the same cannot be guaranteed in strong field regimes or 
at very short or very long scales. In such cases, modifications of Einstein's 
equations could be possible. This is the case for instance in string theory and
higher derivative theories of gravity. In this sense the
equivalence principle, i.e. the description of the gravity field by a metric
tensor, seems to be much more fundamental than the precise form of the 
dynamical equations for the gravitational field.

\subsection{The Cosmological Principle}

Introduced by Einstein in 1917, the term 
\emph{Cosmological Principle} was coined by E. Milne in 1933. 
It states the homogeneity 
(translation invariance) and isotropy (rotation invariance about each point) 
of the three-dimensional space-like slices of the Universe at each instant of 
time. There are very good reasons to formulate this principle. There is
very strong observational evidence that the Universe is isotropic about 
the position ocuppied by the Earth. The distribution of different backgrounds
like galaxies, radio sources, or X-ray background, are pretty isotropic 
around our position on large scales, 
but by far the most precise observational test of isotropy
comes from CMB. After subtracting the dipole term in the angular 
distribution of the CMB temperature on the sky, a 
temperature distribution remains,  whose departure from isotropy  
is only about one part in $10^5$.

From the observation of isotropy around our position, homogeneity
can also be inferred through the so-called \emph{Copernican Principle}.
Copernicus believed that the Sun, not the Earth was the center of the Universe.
Later on, it was discovered that neither the Sun nor the Milky Way occupy 
a special place in the Universe. In 1960 H. Bondi, coined the term 
Copernican Principle to mean that we at Earth do not occupy a special place 
in the Universe. Therefore, since we observe isotropy, it means that isotropy
should be observed from any position in the Universe. Finally, geometry
tells us that isotropy about all points imply homogeneity. 
Thus the observed isotropy of CMB from our position, plus the Copernican 
Principle, imply the homogeneity of the Universe. 

On the other hand recent tridimensional galaxy catalogs 
\cite{2df}, \cite{SDSS}, provide direct observational support for
homogeneity. Figure 1 shows the power spectrum, i.e the density contrast
$\delta \rho/ \rho$ as a function of the averaging scale. 2dF and 
SDSS maps
reach up to 600 Mpc deep in the sky, and from Figure 1 we see that the  
mass density fluctuations in galaxy ditribution diminish to 10\% 
at scales around 400 Mpc. 10\% is not a very high precision 
compared to 
the $10^{-5}$ for isotropy, and so 
the observational check for homogeneity is not so strong as for isotropy.
In addition, it has been also argued \cite{Pietro}, that the
distribution of galaxies could be compatible with a part of a fractal 
up to the limits of the present catalogs. It should be noticed however,
that this does not disprove the Cosmological Principle, nor set reliable
limits to the true scale of homogeneity in the geometry. 
Even if the crossover to homogeneity in luminous matter happens to be 
at a scale which is not yet reached, it should be taken into account that 
luminous matter represents less than 1\% of the energy density 
driving the geometry of the Universe. Nevertheless, understanding 
the relation between the complex galaxy structures and the smooth
CMB represents an extremely interesting and important problem at the heart of 
the theory of structure formation.
\begin{figure}[h]
\centerline{\epsfxsize=8.0 cm \epsfbox{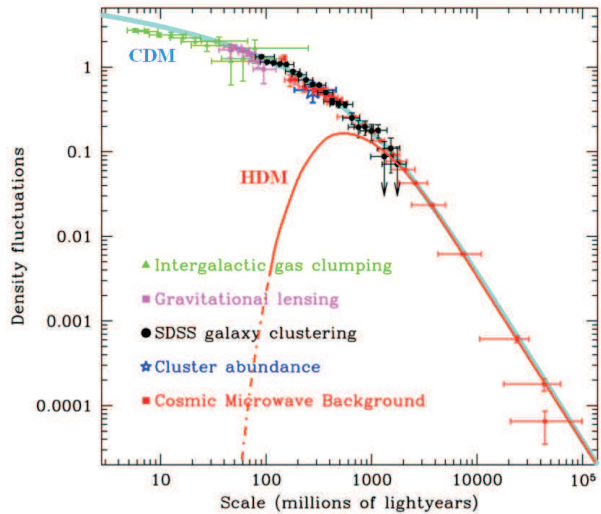}}
\caption{Scale dependence of density fluctuations from CMB anisotropies and large-scale
structure. The continuous line labelled by CDM 
corresponds to a flat model with cold dark matter 
($\Omega_M=0.28$, $h=0.72$) and
a scale invariant spectrum of fluctuations ($n_s=1$). The behaviour for
$\lambda<\lambda_{eq}\simeq 2.9\cdot 10^8$ lightyears is approximately logarithmic, 
whereas it decays as $\lambda^{-2}$ for $\lambda>\lambda_{eq}$. 
The line labelled by HDM represents  the typical behaviour of 
a hot dark matter model, with a sharp cut-off for 
$\lambda<\lambda_{FS}$, (data from SDSS collaboration \cite{SDSS})}
\end{figure}

Another important physical aspect of the cosmological principle is that
homogeneity and isotropy are connected to the fact that the present
Universe comes from a previous state of thermal equilibrium:
the Hot Big Bang. This is a very satisfactory state of affairs
because it makes the Universe comprehensible, since its present state 
can be understood through casual laws without any reference to particular
initial conditions. There is however a caveat. As we shall discuss below,
the existence of a suitable inflationary phase 
guarantees that the region of the Universe that is observable today,
comes from a tiny patch which was causally connected
-therefore able to be in thermal equilibrium- before inflation happened.
Thus, it could well be that we live not very near to the edge 
of a  smooth, homogeneous and isotropic patch caused by inflation,
with the Universe as whole being chaotic. So, it is possible
that the Cosmological Principle is  valid only in a local sense.

The Cosmological Principle together with the Equivalence Principle
dictates the geometry of the Universe given by the Robertson-Walker
metric 

\begin{equation}
ds^2 = - dt^2 + a^2(t) \left( \frac{dr^2}{1-kr^2} + r^2 
(d\theta^2 + {\rm sen}^2 \theta \, d\phi^2 ) \right) \, ~~.
\label{RW}
\end{equation}

In this metric the function $a(t)$ is the scale factor of the Universe, 
the constant $k = 1,0,-1$ specifies the sign of the spatial curvature 
of the Universe, and $t$ is the cosmic time. The cosmic time is the  
time measured by the fundamental or comoving observers which are at rest
with respect to the expansion. They can be characterized as those 
measuring zero dipole anisotropy in CMB. Since the peculiar velocity
of the Sun is roughly 370 km/s with respect to CMB, the time measured by
our watches at Earth coincides with cosmic time with an error of
approximately 1 part in $10^6$. 

The coordinates $r, \theta, \phi $ are (spherical) comoving coordinates 
meaning that comoving observers remain at rest in these coordinates.
Velocities with respect to these coordinate systems are called 
\emph{peculiar velocities}. Galaxies are nearly comoving, their 
peculiar velocities being about a few hundred km/s.

As the Universe expands, physical distances between comoving objects and
wavelenghts scale with $a(t)$. Thus, the wavelength
of a freely propagating photon is stretched in proportion to the expansion
factor from the epoch of emission to detection
\begin{equation}
1+z = \frac{\lambda_{\mathrm{obs}}}{\lambda_{\mathrm{em}}} = 
\frac{a_{\mathrm{obs}}}{a_{\mathrm{em}}} ~~.
\label{redshift}
\end{equation}
This expression also defines the redshift parameter $z$.

The observable region of the Universe at a given time is limited by the 
physical distance that light can travel since the Big Bang.
This distance, called the \emph{particle horizon} is directly related to 
the scale factor by $d_H(t) = a(t) \int_0^t dt'a^{-1}(t') $.

The Cosmological Principle also restricts the form of the material
content of the Universe. Since a perfect fluid can be characterized by 
its isotropy around  observers comoving with the fluid, the 
stress-energy tensor for the material content of the Universe must have the 
perfect fluid form 
\begin{equation}
T_{\mu \nu} = p g_{\mu \nu} + (p+\rho) u_{\mu} u_{\nu}~~, 
\label{perfluid}
\end{equation}
where $p$ and $\rho$ are the pressure and the energy density measured by a
comoving observer, and $u^{\mu}$ is the four velocity of the fluid.

\subsection{FRWL Models}

The Robertson-Walker (\ref{RW}) metric provides the kinematical
framework for cosmological models. It plays a similar role to the metric
of the two-dimensional sphere in the study of Geography. Thus the 
analysis of cosmological observations based only in the RW metric, without
any dynamical assumptions, is sometimes called \emph{Cosmography}.

Cosmological dynamics i.\ e.\  the study of the time variation of
the scale factor and the cosmological densities for the various
matter species entering the composition of the Universe, is obtained 
by the combination of the second basic element of General Relativity:
the dynamical equations for the gravitational field, and the Cosmological
Principle. This leads to the Friedmann-Lema\^{\i}tre equation 
\begin{equation}
H^2 = \frac{8\pi G}{3} \rho + \frac{1}{3}\Lambda - \frac{k}{a^2} ~~, 
\label{FL}
\end{equation}
and the energy conservation equation
\begin{equation}
\frac{d}{dt} (\rho a^3) = -  p \frac{d}{dt} a^3 ~~,
\label{EC}
\end{equation}
where  $H = \dot{a}/a$ is called the \emph{Hubble parameter}. Its present
value $H_0$ is the \emph{Hubble constant}, usually expressed in terms
of the adimensional number $h$ in the form $H_0 = 100 \, h$ km s$^{-1}$ 
Mpc$^{-1}$. Thus, solving Friedmann-Lema\^{\i}tre equation (\ref{FL}) relates 
the expansion rate of the Universe given by the Hubble parameter $H(t)$, 
with the cosmic time $t$ and the red-shift parameter $z$. 

The various species entering the cosmological models are assumed to satisfy
linear equations of state of the form $p = w \rho$. This includes in 
particular the  cases of photons in the CMB or an ultrarelativistic 
plasma $(w=1/3)$, cold non relativistic matter $(w =0)$, and the cosmological
constant $(w=-1)$. In addition, the density parameters for each species 
are defined as $\Omega_i = 8 \pi G \rho_i/3H^2$, where $3H^2/8\pi G$ is 
the \emph{critical density} corresponding to a flat Universe $(k =0)$
with $\Lambda = 0$. Moreover Friedmann-Lema\^{\i}tre equation (\ref{FL}) 
when rewritten as
\begin{equation}
\sum_i \Omega_i + \Omega_{\Lambda} - \frac{k}{a^2 H^2} = 1 ~~,
\label{Fdens}
\end{equation}
relates the density parameters to the spatial curvature, including
$\Omega_{\Lambda} = \Lambda/3H^2$  as the density parameter for the 
cosmological constant. Thus flat universes are those fulfilling the condition 
$\sum_i \Omega_i + \Omega_{\Lambda} = 1$. It is important to remark 
that all universes having a Big Bang, 
are nearly flat at early times since the density parameter for curvature
$\Omega_k = -k/a^2 H^2 \rightarrow 0$ as $t\rightarrow 0$

On the other hand the temporal evolution for the densities is given, in terms
of the red-shift parameter $z$, by the energy conservation equation 
(\ref {EC}), and the equation of state
\begin{equation}
\displaystyle{\Omega_i H^2 (1+z)^{\left( -3-3w_i \right)} = 
\frac{8\pi G}{3} \rho_i (1+z)^{\left( -3-3w_i \right)} = \mathrm{const.} } 
\label{evomega}
\end{equation}
 
As we shall discuss below, observations favour a present Universe
which is nearly flat and  composed of cold matter and a cosmological
constant, with present values of density parameters 
$\Omega_M \approx 0.27$ and  
$\Omega_{\Lambda} \approx 0.73$. As shown in Figure 2 the selection 
of a region in the $\Omega_M$-$\Omega_{\Lambda}$ plane centered about
these values, arises from the combination of observational information 
from supernova type Ia Hubble
diagrams, CMB anisotropies, and clustering of galaxies in LSS.  
Taking into account the measured value of the Hubble constant
$h \approx 0.72 (10\%)$ \cite{Hubble} and solving Friedmann-Lema\^{\i}tre
equation for these models, yields an age for the Universe around 
13.7 Gyr, and a distance to the particle horizon about 14-15 Gpc. 

According to the evolution equation for the densities (\ref{evomega}),
$\Omega_{\Lambda}$ decreases and $\Omega_M$ increases with increasing 
red shift. Thus, going backwards in time, for $t \lesssim 1$ Gyr, 
$\Omega_{\Lambda}$ becomes negligible and we are left with a critical 
Universe dominated by cold matter, until radiation begins to dominate.
In addition, from the analysis of CMB anisotropies,
and from BBN, it follows that only about $15\%$ of this cold matter
are baryons, the rest being \emph{cold dark matter}.

Planck's formula for blackbody radiation translates the present CMB temperature
$T = 2.725$, into an energy density of CMB photons 
$\Omega_{\gamma} h^2 = 2.48 \times 10^{-5}$. Including three massless
(or very light) species of neutrinos, the total energy density in radiation 
would be at present $\Omega_R h^2 = 2.48 \times 10^{-5}$, which is negligible
in front of the energy density in cold matter. However as we go 
backwards in time, the number density of radiation particles 
scales as $a^{-3}$,
and the wavelength shrinks in proportion to the scale factor $a$. Therefore,
the energy density in radiation scales as $a^{-4}$ while the energy
density in cold matter goes with $a^{-3}$. As a consequence, at sufficient
earlier times, radiation dominates the energy density of the Universe.
For $h \approx 0.72$ and $\Omega_M \approx 0.27$ the energy densities in
matter and radiation become equal at $z_{\mathrm{eq}} \approx 3300$, 
which corresponds to an approximately  55 kyr old Universe.

\section{Hubble Law, Hubble Diagrams and Dark Energy}

In the approximation in which galaxies are comoving, the physical distance
to a given galaxy scales with $a(t)$, and consequently its recession
velocity $V$ is related to its physical distance $d$ at a given time, by  
\begin{equation}
V = H d ~~.
\label{THL}
\end{equation} 
This is the \emph{theoretical Hubble law} which is exact, and a direct
consequence of the Robertson-Walker form of the cosmic metric.
This relation cannot be directly checked because neither the recession
velocities nor the physical emission distances to galaxies are empirically
measurable.  

The Robertson-Walker form of the metric was stablished in 1936. Earlier,
in 1929 E. Hubble found the \emph{empirical Hubble law}
\begin{equation}
z = H_0 d_{\mathrm{L}} ~~,
\label{ehl}
\end{equation} 
linearly relating the red-shift of galaxies to their luminosity distance.
The luminosity distance is defined as $d_{\mathrm{L}} = 
\sqrt{L/4 \pi \mathcal{F}}$, where $L$ is the absolute luminosity
of the source and $\mathcal{F}$ its apparent luminosity, i.\ e.\ the
flux of energy received in the collecting surface of the telescope. 
So, luminosity distance $d_{\mathrm{L}}$  is defined as such that a source 
of absolute luminosity $L$, located in a static Euclidean space, would produce
a flux $\mathcal{F}$ at distance $d_{\mathrm{L}}$. From RW metric, it 
follows that the relation between $d_L$ and the red-shift parameter $z$
is nonlinear. To second order this relation takes the form
\begin{equation}
H_0 \, d_{\mathrm{L}}(z) = z + \frac{1}{2} (1-q_0) z^2 + \cdots 
\label{decpar} ~~,
\end{equation}
where $q = - a \ddot{a}/\dot{a}^2 $ is the \emph{deceleration parameter} of 
the Universe and $q_0$ its present value. It follows then from RW metric,
i. e. from Cosmological Principle, that the empirical
Hubble law can be expected to be true only for $z \ll 1$.
On the other hand for  $z \ll 1$, the approximate equalities
$d \approx d_{\mathrm{L}}(z) $ and $ V \approx z $ hold. Thus empirical and 
theoretical Hubble laws coincide in this regime. In this weak sense checking 
Hubble's law is also a check of the RW metric.

To accurately check Hubble's law, measuring $H_0$ within the linear
approximation to (\ref{decpar}), and eventually going
deeper in red-shift to determine $q_0$, has been a central research
program in Cosmology since 1929, called Hubble program. The key 
observational tool for this endeavour are \emph{standard candles}:
luminous sources whose absolute luminosity has been properly calibrated.
Once a class of sources has been calibrated, a Hubble diagram can 
be obtained by representing these sources in a two-dimensional
plot of luminosity distances versus red-shifts, the final goal being
to extract from the observational points the cosmological parameters
$H_0$ and $q_0$. This program received a major boost with the launching
of the \emph{Hubble Space Telescope} (HST) in 1990, whose so-called 
\emph{Key project} was to perform a precise measure of $H_0$.  

Measuring cosmic distances starts by the trigonometric parallax
of nearby stars, due to the annual motion of the Earth around the Sun.
From this starting point, a \emph{cosmic distance ladder} is built by means
of standard candles. This method works typically by finding precise 
correlations between the absolute luminosity and another observable
for a definite class of objects. The first and basic step 
is provided by Cepheid variable stars, 
whose absolute luminosity is tightly related to its period.
HST has been able to resolve thousands of Cepheid 
variables in galaxies up to 20 Mpc. Once the distances to these nearby 
galaxies are
fixed, five different methods are used to go up to 400 Mpc.
Three of them are based on global properties of spiral and elliptical 
galaxies: Tully-Fisher relation that links rotation velocities of spiral
galaxies to their luminosity, relation  between star velocities dispersion
and luminosity in ellipticals, and fluctuations in galaxies surface 
brightness. The other two are based on the use of supernovae type Ia (SNe Ia),
and supernovae type II (SNe II) as standard candles. 
The combination of all these methods yields for the
Hubble constant a weighted average $H_0 = 72 \pm 8$ km s$^{-1}$ Mpc$^{-1}$
\cite{Hubble}. It is most remarkable that this value for $H_0$ agrees
with the value obtained by the analysis of the CMB anisotropy map
provided by WMAP: $H_0 = 71  $ km s$^{-1}$ Mpc$^{-1}$ $(5\%)$.

Due to their big absolute luminosity, SNe Ia are 
the most far reaching standard candles in red-shift. This
makes these objects ideal tools to investigate
the deceleration parameter $q_0$.
In 1998 two groups: the \emph{High z Supernovae Search Team} (HZT) and the
\emph{Supernova Cosmology Project} (SCP) using SNe Ia,
reported the discovery of the accelerated expansion
 of the Universe ($q_0 < 0$) \cite{SN}. 
The result appears through an exceeding faintness of supernovae 
as they would have in a decelerating universe. Notice that from  eq. 
(\ref{decpar}), an accelerating universe ($q_0 < 0$) results in
bigger luninosity distances -hence, fainter objects- for the
same red-shift than a decelerating one ($q_0 > 0$).
Thus, once other astrophysical effects like dust or evolutionary effects
have been discarded, the exceeding faintness of supernovae should be
interpreted as due to the acceleration of the Universe expansion. 

Friedmann's equations (\ref{FL}), (\ref{EC}) imply that an
accelerating Universe should contain part of its energy density
in a substance with equation of state parameter  $w < -1/3$. This
kind of substance goes under the name of \emph{dark energy}, and the 
most obvious candidate is a cosmological constant ($w = -1$).
For a model composed of dark matter plus a cosmological 
constant, the deceleration parameter 
$q = \frac{1}{2} \, \Omega_M - \Omega_{\Lambda}$,
and as shown in Figure 2, the supernovae data select a linearly
shaped maximum likelihood region in the $ \Omega_M - \Omega_{\Lambda}$
plane. So, supernovae data alone, suffice to stablish the acceleration of
the Universe with a very high confidence level. When supplemented with
the data coming from CMB and LSS, a best fit is obtained for
$\Omega_M \approx 0.28$ and $\Omega_{\Lambda} \approx 0.72$, corresponding
to a nearly flat Universe. 

Although the LSS models favored the so-called $\Lambda$CDM (cosmological
constant plus cold dark matter) scenario, the discovery of the acceleration of the 
Universe in 1998 came as a surprise, since a universe filled with
cold matter ($w = 0$) decelerates ($ q = \frac{1}{2} \, \Omega_M > 0$). However
as explained above, when we go backwards in time,  $\Omega_M$ grows and
$\Omega_{\Lambda}$ decreases. So, at earlier times there should
have been a decelerating period of the Universe. In fact,
very recently \cite{vuelta}, using the HST, 16 new type Ia supernovae
has been found at very high red-shift, up to $z\approx 1.7$, which 
give conclusive evidence for this decelerating period. These newly
discovered SNe Ia, together with the 170 previously reported
confirm the concordance model with 
$\Omega_M \approx 0.28$ and $\Omega_{\Lambda} \approx 0.72$,  
and give a value for the red-shift of the transition between the accelerating
and deccelerating epochs $z = 0.46 \pm 0.13$. It is the most remarkable
that the confirmation of the effect of deceleration in supernovae, almost
fully rule out alternative astrophysical explanations like dust or 
evolutionary effects for the luminosity distance versus red-shift distribution,
since it is very unlikely for these effects to exactly mimick the
deceleration/acceleration transition.

\begin{figure}[h]
\centerline{\epsfxsize=7.0 cm \epsfbox{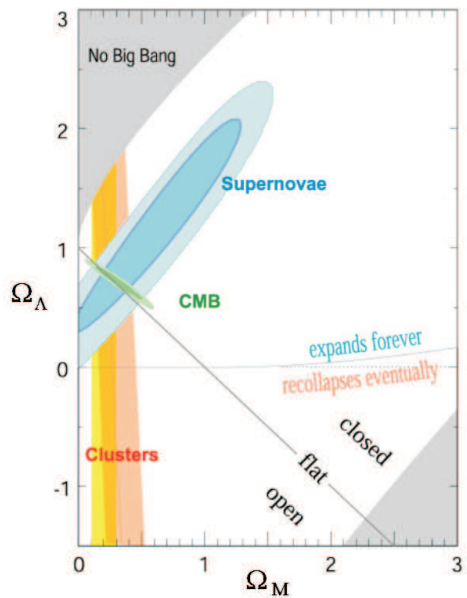}}
\caption{Matter density $\Omega_M$ vs. vacuum energy 
density $\Omega_\Lambda$ 68$\%$ and 95$\%$ C.L.
contours for supernovae, cluster and CMB data,  
(from Supernova/Acceleration Probe: SNAP collaboration \cite{SNAP})}
\end{figure}

In addition to the cosmological constant, other forms of dark energy 
could be possible with a different or even a variable equation of state $w$.
An interesting class of models are \emph{quintessence} models in which dark
energy is the energy density of an evolving scalar field, much the same
way as the inflaton during the inflationary phase of the Universe. 
For a constant equation of state, SNe Ia data yield 
$w= -1.02_{0.19}^{0.13}$ and $w < -0.76$ to
95\% confidence level \cite{vuelta}.
An alternative proposal to \emph{dark energy} as an explanation for
the deceleration/acceleration of the Universe could be 
the weakening of 
gravity in our 3+1 dimensions by leaking into the extra dimensions
as suggested by string theories. 
A major observational effort will be 
needed to discriminate among these competing models.

\section{Large Scale Structure and Dark Matter}
Far from a smooth distribution, 
 matter exhibits a complex clustering pattern in the Universe. Thus, 
there are regions in which 
matter is strongly clumped forming galaxies, clusters and 
even larger structures, whereas at the same time, 
we can also find almost empty regions with very low densities.
In fact,  strong inhomogeneities
can be found at
galactic scales ($\sim$ 10 kpc), where the density
contrast can be as large as $\delta \rho/\rho\sim 10^2$. 
Galaxies, which can be considered as the
elementary building blocks of structures,  
are not homogeneously
distributed either, but grouped
hierarchically into  groups, clusters and superclusters, the latter
ones extending over distances of tens of Mpc. Filament-like chains
of galaxies connect different superclusters in a network with
scales of around 100 Mpc. Most of the matter distributes on the walls
of this cell-like structure with large voids in between.
 In Figure 1
we can see a data plot showing the scale dependence of the density 
contrast. 
Each point represents the density fluctuation measured at 
a given scale $\lambda$, which is 
obtained by comparing the avarage density within a sphere of
radius $\lambda$, as it is placed at different spatial positions.
As we see from the data, the density contrast declines as we
take larger and larger spheres, in agreement with the Cosmological
Principle. Thus, we can conclude that the Universe can be
considered as approximately 
homogeneous only on very large scales ($\sim$ 1000 Mpc).

The SCM allows us to understand the growth
of these structures  from seeds
of primordial  density fluctuations. However,  
the origin of such seeds
is left unspecified,  this being one of the most important 
limitations of the classical standard cosmology. 
In Section 7 we will see  a possible
generation mechanism based on the idea of inflation. Leaving aside
the issue of the origin,  the growth mainly 
takes place during the matter dominated
era as more and more matter 
is attracted towards the initially  overdense regions. 
When the structure is
sufficiently large (larger than the so-called Jeans scale), 
its gravitational self-attraction is able to decouple
it from the Hubble expansion, forming a bound object. For small
density fluctuations (linear regime), the growth rate
is linear with the scale factor. This means that the 
total growth from the matter-radiation decoupling time 
($a_{dec}\sim 10^{-3}$) until present ($a_0=1$) 
 would be  a factor $10^3$.
However, the amplitude of CMB fluctuations at that time, measured
by the COsmic Background Explorer (COBE) satellite, was only $\delta \rho/\rho\sim\delta T/T\sim 
 10^{-5}$. This obviously poses a problem since
fluctuations have not had enough time to reach the non-linear regime
$\delta \rho/\rho \gg 1$. Notice that in the previous reasoning it is 
assumed that matter fluctuations are comparable to temperature 
fluctuations at decoupling. This is indeed the case for 
baryons, which were coupled
to photons until decoupling time, and implies that
a universe dominated by baryons at the time of
decoupling is not consistent with galaxy formation.

However, if there existed a new
type of weakly coupled matter, which did not interact with 
radiation, its density fluctuations could have started growing
much before, thus explaining the apparent
mismatch. This is one of the strongest arguments in favor of
the existence of dark matter. Once, baryons and radiation decouple,
baryons will fall in the potential wells created by dark matter, 
their density fluctuations acquiring the same amplitude as that
of dark matter. 

The nature of dark
matter determines the final density distribution at different
scales. At present  there are important projects which aim to
collect information about the distribution of galaxies in the Universe.
They are galaxy redshifts catalogues such as the 2dFGRS  \cite{2df} 
and the SDSS  \cite{SDSS}. 
The first one, which has
been completed recently, has measured the redshift of 221000 galaxies
over a five years period. The SDSS is in progress and is expected
to measure the position and the absolute brightness of 100 million
celestial objects. The information obtained from these
catalogues combined with that coming from CMB anisotropies 
and high-redshift supernovae observations is allowing us to 
determine the 
cosmological parameters with unprecedented accuracy (see Fig. 2), 
and to shed light
on the nature of dark matter.  

Let us see how dark matter
affects structure formation. For that puropse it is convenient to 
differentiate
between the so-called hot (HDM) and cold dark matter (CDM) 
scenarios.
In the hot case,  dark matter is made out of light particles, 
which were still ultrarelativistic  at the beginning of the 
structure formation period, with the typical candidate being  
a light neutrino. Since hot dark matter particles propagate 
 close to the speed of light, they can escape from overdense
regions into underdense ones,  erasing the density fluctuations
on scales smaller than the free-streaming scale $\lambda_{FS}$. 
This is the maximum 
distance that a particle can travel from the initial time 
until matter-radiation equality. Typical values are $\lambda_{FS}
\sim 40$ Mpc, corresponding to the size of a large cluster, 
for a neutrino (or any other light particle) 
mass around $m_\nu\sim 30$ eV. This means that, in
this scenario, galaxies cannot grow directly from 
 primordial fluctuations. Superclusters  should form 
first, reach 
the non-linear regime and then, by fragmentation, 
give rise to small clusters and galaxies. 
As a consequence  the power spectrum of density fluctuations
should be strongly peaked around the $\lambda_{FS}$ scale (see Fig. 1).
However, recent CMB data together with 2dFGRS or SDSS 
information on the matter
power spectrum, constrain the above effect.
The results show that hot dark matter cannot be the dominant dark matter
component and $\Omega_{\nu}h^2\leq 0.0076$ (2dF+WMAP at 95 $\%$ C.L.). 
This bound can be
translated into a very strict limit on the sum of the
neutrino masses $\Sigma_i\; m_{\nu_i}\leq 0.71$ eV, which improves by several
orders of magnitude the laboratory limits. 

In the cold dark matter case, dark matter particles are already
non-relativistic at matter-radiation equality. Thus, free-streaming
damping is not a problem. 
As shown in Fig. 1, an enormous variety of observations at very large scales 
($\gsim 1$ Mpc), from  cosmic
microwave background anisotropies, galaxy surveys, cluster abundances 
or Ly-$\alpha$
forest  are successfully explained within the CDM framework. 
Despite its success  at large scales, the model exhibits certain 
difficulties at sub-galactic scales. In particular, high resolution 
N-body simulations
of dark halos show cuspy density profiles  which contradict 
observations from low
surface brightness galaxies  and dwarfs   
which indicate flatter density profiles. In addition,
CDM also predicts too many small subhalos within  
simulated larger systems, in 
contradiction with observations of the number of 
satellite galaxies in the Local
Group.  In any case a spatially flat Universe with 
cosmological constant and  cold dark matter ($\Lambda$CDM) is
generally accepted 
at present as the  Standard Model in Cosmology.  
  
\subsection{Nature of Dark Matter}

Apart from the difficulties of a baryon dominated universe to 
explain galaxy formation, 
there are additional evidences that the luminous mass of the Universe 
is only a small fraction of the total matter density \cite{DM}.  This deficit 
is present in two different contexts: first at galactic scales, where  
dark matter is believed to form  spherical  halos several times
bigger than the galactic disks; and second at cosmological scales.   
Many different kinds of observations such as rotation
curves of galaxies, weak gravitational lensing, cluster abundance,  
virial motions in clusters, matter power spectrum, CMB anisotropies, 
...,  agree in a value for the 
total matter $\Omega_M= 0.27\pm 0.04$. This value should be
compared to the baryon density $\Omega_{B}=0.044\pm 0.004$ and 
to the luminous mass density $\Omega_{lum}= 0.006\pm 0.003$, i.e
we find $\Omega_{lum}<\Omega_{B}<\Omega_{M}$. We thus have 
two dark matter
problems, namely, there are missing baryons which do not contribute to
the luminous matter and there is non-baryonic dark matter
which make up most of the matter of the Universe. 

Concerning the baryonic dark matter
problem, it is difficult to find dark baryons in the galactic halos. 
They could be present in the form of hot
or cold clouds of hydrogen, although an entire
halo made of gas would conflict with observations of
absortion or emission of radiation. They could  form massive compact 
halo objects (MACHOs) similar to big planets. However 
the current limits 
from EROS and 
MACHO microlensing observations show that less than 25$\%$ of 
standard halos can be composed of MACHOs with masses between
$10^{-7}-1$ M$_\odot$. 
    
On the other hand, the nature of non-baryonic dark matter 
is even a greater mistery. Different possible explanations
include: massive neutrinos, modifications of gravity at large
distances or the existence of a background of new weakly interacting
massive particles (WIMPs). A natural dark matter particle should be
 neutral, stable, massive and weakly interacting, so that its relic
density could contribute in an important way to the matter density. 
Accordingly a massive neutrino would be the most 
economical solution. However, different limits prevent neutrinos
from being a viable candidate. Thus, apart from the strong limit 
coming from 2dF + WMAP mentioned above, simply imposing that relic neutrinos
do not  
overclose the Universe, i.e.  $\Omega_\nu\lsim 0.3$, then 
their masses should be either smaller than
$\sim 20$ eV or larger than  $\sim 20$ GeV. 

Let us study in  more detail each
possibility. In the case of light (hot) neutrinos, if
they are required to make up the galactic halos, their mass density
should be $\rho_{halo}\simeq 0.3$ GeV cm$^{-3}$. However, their 
number density cannot exceed  the limit imposed by the Pauli 
exclusion principle, so that  their masses should be 
sufficiently high. In particular we get $m_\nu\gsim 20$ eV for
spiral galaxies and  $m_\nu\gsim 100$ eV for dwarf galaxies 
(Tremaine-Gunn limit), in contradiction with the overclose limit.
In the case of heavy (cold) neutrinos, the overclose limit is 
much larger than the laboratory limits on the three known
neutrino species, but still there is the possibility of the
existence of a stable heavy fourth generation of Dirac or Majorana 
neutrinos. Current direct
detection experiments have enough sensitivity to detect
halo particles with cross-sections typical of weak interactions 
and masses above 20 GeV. However, at present there is 
no compelling evidence of the detection of 
such particles, so that 
a fourth generation of neutrinos is excluded. 
(DAMA experiment claims the detection of an annual modulation
in its dark matter signal. However, such a result seems to be 
incompatible
with other direct detection experiments as CDMS). 

The absence of cold dark matter candidates within the 
known particles is 
one of the most pressing arguments for the existence of
new  physics beyond the Standard Model, either as new particles
or as modifications of the gravitational interaction at large
distances. 
Among the proposed new particle candidates, we find, on one hand,  the axion which 
is the Goldstone
boson associated to the spontaneous breaking of the Peccei-Quinn 
symmetry postulated
to solve the strong CP problem of QCD. The production  of axions in 
the early
Universe  mainly takes place through the so-called misalignment 
mechanism in which
the $\Theta$ angle is initially displaced from its
equilibrium value  $\Theta=0$, and oscillates coherently. Such 
oscillations can be
intrepreted as a zero-momentum Bose-Einstein condensate which 
essentially behaves
as a non-relativistic matter fluid. Despite the fact that axions 
are light
particles, this non-thermal 
mechanism produces  cosmologically important energy densities. 
 On the other hand we have the thermal relics, produced by the
well-known freeze-out mechanism in an expanding Universe. 
They are typically 
weakly interacting massive
particles (WIMPs) such as the neutralino in supersymmetric 
theories
(for a 
recent review see  \cite{DM}).
In addition to their weak interactions with ordinary particles 
included in the EPSM, 
these candidates
usually have also very weak self-interactions (collisionless).

At present there are several kinds of experiments (ground based and 
satellite borne) which aim to detect cold dark matter halo particles, either
directly or indirectly. Direct detection experiments are based
on the possibility of measuring the recoil energy that a target 
nucleus acquires in the elastic collision with a DM particle.
Some of the experiments in progress are: DAMA, CRESST and GENIUS, 
at Gran Sasso Laboratory or CDMS at Soudan mine. The indirect 
experiments are based
on the possibility of detecting the annihilation products of halo
DM particles. Typically they include: gamma ray telescopes 
such as MAGIC (ground based) or GLAST (satellite)
which could be sensitive to annihilations into pairs of photons; 
antimatter detectors such as AMS which can detect positrons produced
in $e^+e^-$ annihilations; and finally high-energy
neutrino telescopes such as ANTARES or AMANDA which will be sensitive
to neutrino-antineutrino annihilations.
The projected sensitivity of these experiments covers a part of
the  parameter regions (masses and interaction cross-sections)
which will be explored by future particle accelerators
such as LHC or Tevatron II. However their  
relative low cost make them  very promising alternatives for  
finding  new physics.

\section{The Cosmic Microwave Background}
The existence of a Cosmic Microwave Background with present temperature 
around 5K, was theoretically predicted in 1948 by
G. Gamow, R. Alpher and R. Hermann, as a necessary relic of a hot phase
of the Universe in which the light elements should have been cooked
through nuclear reactions starting with primordial protons. This 
work largely ignored during almost two decades, can be regarded with
hindsight as the foundational paper of the Hot Big Bang model.  
On the other hand, in year 1964, A. Penzias and W. Wilson who where working at 
Bell Telephone Laboratories to fit an antenna for satellite communications
operating in the microwave range, found to his annoyance an ``excess"
radio noise isotropically distributed and corresponding to a temperature
of about 3K, whose origin they did not know and did not hypothesized about.  
In the same year the group of theoreticians of Princeton University:
B. Dicke, P. Peebles, P. Roll, and D. Wilkinson, who did not know or 
had forgotten about the work of Gamow et al, where following a similar 
line of reasoning. In fact they  where thinking about building a radiometer 
to detect the fossil radiation left out from a primitive dense and 
hot phase of the Universe, when they knew about Penzias and Wilson 
observations and correctly interpreted them as the cosmic background radiation
they were looking for. Similar considerations were 
also done by Y. Zeldovich and his group in Moscow around the same time.
Thus CMB was accidentally discovered in 1964, and Penzias
and Wilson (but not Gamow et al. nor Dicke et al.) were awarded 
the Nobel prize in 1978.

Since the temperature of the photons in CMB scales according to 
$T \propto a^{-1}$, the existence of CMB together with the expansion
of the Universe imply a hot early phase, which becomes hotter and denser 
as we go backwards in time. In this 
primitive epoch, the Universe is a plasma containing more and more
species of particles as we go backwards in time and new channels are opened
for pair production to the increasingly energetic photons. 
In this way, when the photons are cold enough, $T \sim 3000$ K,
the baryons and electrons \emph{recombine} to form neutral hydrogen and 
helium atoms, and the photons are free to propagate: the Universe becomes 
transparent. This happens for a red-shift parameter 
$z_{\mathrm{rec}} \sim 1100$, which corresponds to an age of about 380 Kyr. 
The recombination red-shift $z_{\mathrm{rec}}$ defines a \emph{last 
scattering surface} for charged particles and photons, or cosmic
photosphere, where CMB is coming from. Indeed, this is not a mathematical 
surface but it has a thickness which can be modeled by a Gaussian 
visibility function with a width $\delta z \simeq 80$.

CMB radiation is therefore a relic from $z_{\mathrm{rec}} \sim 1100$, beyond 
which the Universe is optically thick in almost all wave bands, and it carries
vital information about the processes and features of the 
early Universe \cite{CMB}. 
So, since its discovery and specially in the last fifteen years there 
has been an intense observational effort on CMB. The COBE  satellite launched in November 1989 made a major breakthrough. 
With the FIRAS (far infrared absolute spectrophotometer) instrument on board, 
it was stablished the almost perfect blackbody spectrum of CMB, 
which witnesses the perfect thermal equilibrium state from which our Universe 
is coming from. It is most remarkable that such a pure blackbody 
spectrum has never been observed in laboratory experiments. 
In addition, with the help of the DMR (differential microwave radiometer) 
instrument, a map of CMB temperature 
anisotropies was obtained for the first time. After COBE a series of ground and balloon based 
measurements: ARCHEOPS, BOOMERANG, DASI, MAXIMA, VSA and for smaller scales 
CBI and ACBAR, have been carried out to improve the quality of temperature 
anisotropies distribution data. The most important recent advance has been 
the fisrt year of operation results from NASA's WMAP (Wilkinson Microwave 
Anisotropy Probe) \cite{CMB,WMAP}. Launched in June 2001, the first data release in 
February 2003, corresponds to a twofold full coverage of the sky and 
provides a
much more precise anisotropies map than COBE's (see Fig. 3). WMAP is funded
to operate for at least  another three years, completing and eightfold covering of 
the sky and increasing statistical accuracy.  After that, ESA's satellite 
Planck \cite{Planck}, scheduled for launch in 2007, will take over. As WMAP, Planck 
will be stationed at Lagrange L2 point of the Sun-Earth system, and will
cover the full sky reaching an angular resolution up to one tenth of degree.
    
\begin{figure}[h]
\centerline{\epsfxsize=10.0 cm \epsfbox{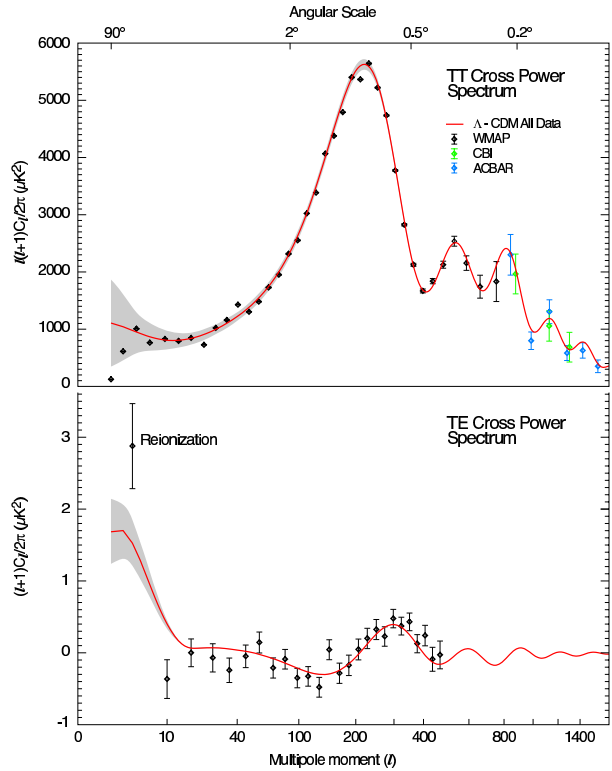}}
\vspace{-2cm}
\caption{TT and TE angular power spectra from WMAP one-year data,
\cite{CMB} }
\end{figure}

The CMB temperature distribution in the sky, being a function defined
on a sphere is most naturally analyzed through an spherical harmonics
expansion
\begin{equation}
T(\theta,\phi) = \sum_{\ell m} a_{\ell m} Y_{\ell m}(\theta, \phi)
\label{spherical}
\end{equation}
The monopole component gives the mean temperature of CMB $T = 2.725 \pm
0.001$ K, which according to Planck's law for blackbody radiation, 
corresponds to a photon number density $n_{\gamma} = 411$ cm$^{-3}$, 
and an energy density $\rho_{\gamma} = 0.260$ eV cm$^{-3}$. The largest 
anisotropy is the $\ell = 1$ dipole term with amplitude $3.346 \pm 0.017$ mK
interpreted as the result of the Doppler shift caused by the Solar system
motion relative to the CMB. The implied velocity for the Sun is $v = 368 
\pm 2$ km s$^{-1}$, directed towards Hydra-Centaurus (more
precisely, right ascension $= 166 \pm 3$ and declination $= -27.1 \pm 3)$. 
This interpretation is reinforced by the yearly modulation of the anisotropy due to
Earth's motion around the Sun, and also by measurements of the
velocity field of local galaxies. This rather impressive finding about what 
could be called ``our home's absolute velocity" embodies a delicious irony: 
the negative result of the Michelson-Morley experiment in 1887 in detecting 
Earth's motion with respect to aether (the hypothetical mechanical 
medium supporting the propagation of light), motivated the introduction
of Special Relativity. On these foundations General Relativity was built
and Cosmology developed during the XXth century. Finally, towards the end of
the last century, the peculiar velocity of the Earth was measured with 
light (CMB) playing the role of aether, and without contradicting Relativity.
Simply, CMB photons materialize a convenient comoving coordinate system.

Once the monopole and the dipole have been removed from the expansion 
(\ref{spherical}), we are left with the CMB intrinsic anisotropies which 
are of the order of, or below  $10^{-5}$ in all angular scales, and contain 
the imprints of the early Universe physics at radiation-matter decoupling.
Most of the cosmological information is contained in the two point
temperature-temperature (TT) correlation function. 
This quantity is defined by
averaging the product of the fractional temperature deviations in 
directions $\vec{n}$ and $\vec{n'}$ over the sky, and expanding the result 
in Legendre polynomials
\begin{equation}
C(\theta) \equiv \langle\,  \frac{\Delta T (\vec{n})}{T} 
\frac{\Delta T (\vec{n'})}{T} \, \rangle = \sum_{\ell = 0}^{\infty} 
\frac{2\ell + 1}{4 \pi} C_l P_l (\cos \theta)
\label{twopoint}
\end{equation}
The expansion coefficients $C_l$, when represented as function of $\ell$
(more suitably $\log \ell)$ give the so-called \emph{angular power spectrum}
which is the key function in comparing theory and observations. 
The cosmological parameters affect the form of this function and this 
is the way in which they can be deduced from CMB observations. 
Several physical mechanisms contribute to the angular power spectrum at
different angular scales $\theta \sim \pi/\ell$. 
An important milestone is the size of the comoving Hubble radius at 
decoupling. This is half the size of the comoving  particle horizon at that 
time if inflation had not happened before, and far from covering the whole sky,
it subtends an angular opening $\theta \sim 0.86 \sqrt{\Omega}$, 
with $\Omega$ being the present total density parameter of the Universe.  
This is the origin of the horizon problem that inflation solves as we 
will discuss below.   

The description of the physics contained in $C_l$, can be separated
into three main regions. i) \emph{The Sachs-Wolfe plateau} for $\ell \leq 100$:
This region corresponds to angular scales bigger than the Hubble radius at
decoupling and it is dominated by the so-called Sachs-Wolfe effect: the 
photons coming from denser regions have to climb out of deeper gravitational
potential wells and become redder. A nearly scale invariant spectrum
of density perturbations, as predicted by inflation, agrees with a plateau
for this effect. Also, the integrated Sachs-Wolfe effect due to the time 
variation of the gravitational potential along CMB photons world lines, 
and gravity waves too, are expected to contribute in this large angular
scale region, although these effects are buried in the \emph{cosmic variance}.
This means that doing an statistical analysis of fluctuations for a given set 
of cosmological parameters, but having only one realization of these 
fluctuations (our Universe) forces to introduce a kind of ergodic hypothesis:
averages of patterns over parts of the sky, extended to the full sky by 
periodicity, would be equal to averages over different full sky realizations 
of perturbations for the same set of parameters. Indeed this method becomes 
increasingly uncertain at large angular scales and this is the meaning 
of the term cosmic variance.    

ii) \emph{The acoustic peaks} for $ 100 \leq \ell \leq 1000$. Before 
decoupling, photons, electrons and baryons form a tightly coupled fluid
with the photons providing the pressure and the baryons the inertia.
The fluid supports acoustic waves whose fundamental halfperiod is determined
by the Hubble length at decoupling, and in turn its wavelength is obtained 
from the known value of the sound speed in the plasma. In this oscillating 
plasma  denser regions correspond also to hotter regions (notice the opposite
sign of this effect as compared to Sachs-Wolfe) according to Planck's law.
Therefore a series of peaks in $C_l$ are expected. The first one 
corresponding to the fundamental acoustic mode, and subsequent ones for
higher harmonics. These peaks were theoretically predicted by P. Peebles, and
Y. Zeldovich and collaborators already in 1970, and the empirical proof of 
their existence is a major success of modern cosmology. The position and 
height of the acoustic peaks encode information about the cosmological 
parameters. Indeed the angular scale subtended today by the fundamental 
acoustic wavelength depends on the underlying geometry. This is how 
the position of the first peak, stablished by WMAP to be around $\ell \sim 220$,  
results in a flat Universe with $\Omega = 1.02 \pm 0.02$.  
This way of stablishing the flatness of our Universe is in fact very similar
in essence to the method attributed to Gauss, who supposedly 
measured the three angles formed by three peaks in the Harz mountains, 
in order to check space geometry. In this case the triangle is the
one formed by the acoustic fundamental halfwavelenght sitting on the last 
scattering surface, and two lines of sight from its extremes to us.
Thus in a $\Lambda$CDM model, the position of the first acoustic peak
determines $\Omega_M + \Omega_\Lambda$, while the difference 
$\Omega_M - \Omega_\Lambda$ can be extracted from SNe Ia Hubble diagrams
as explained above. The second peak is not as high as the first one
because the baryons feel the radiation pressure but cold dark matter does not,
and as a consequence the relative height of the first and second peak
gives the amount of baryons in the Universe. Combined results from 
WMAP, CBI and ACBAR yield $\Omega_B h^2 = 0.023 \pm 0.001$. 
In turn this result, together with the photon number density, fixes a
very important cosmic number, namely the  baryon/photon ratio or specific
entropy of the Universe, only from CMB data. The resulting value is
$\eta_{10} \equiv 10^{10} \eta = 6.14 \pm 0.25$. It is very remarkable,
and a strong indication of the maturity of the SCM 
that this value is consistent with the determination of $\eta$ via 
the physics of BBN when the Universe was three minutes old. 
In addition, the relative height of the first three peaks provides
a determination of $\Omega_M$. The Hubble parameter $H_0$ can
be also extracted from the angular power spectrum, although the dependence 
of its shape with respect to $H_0$ is more involved. WMAP gives
$h = 0.71 (5\%)$, also in good agreement with Hubble diagrams.

iii) \emph{The damping tail} for $\ell \geq 1000$. As stated above
the transition to transparence is not instantaneous and the last 
scattering surface has a thickness. This leads to the so-called
\emph{Silk damping} of the anisotropies for angular scales smaller
than this thickness. In addition, gravitational lensing by non-linear
structures at low red-shift, like clusters of galaxies, also deform
and smooth the primordial angular spectrum at small scales. As 
a consequence there are not much primordial anisotropies to observe 
below 5' of arc.

Electron-photon Thomson scattering at the last scattering surface transforms
anisotropies into CMB photons polarization. The analysis of polarization 
leads to four new non-vanishing two sky points correlators, with their 
corresponding angular spectra. The theoretical and observational analysis
of these spectra lies at the present frontier of CMB research. For example,
Planck satellite is expected to do a significant advance in this respect.
In particular the gravity wave contribution to CMB anisotropies could
be observed. If the gravity wave perturbations were produced by 
inflation, these observations would determine the energy scale at which 
inflation happened. Also the WMAP results concerning polarization measurements 
have recently uncovered an earlier than expected reionization of the 
Universe  at red-shift $z$ around 20 
(see the TE power spectrum in Figure 3). 
This means that the first stars in the 
Universe were formed as early as a few hundred million years after 
Big Bang. Such an early star formation could be challenging for the
theory of inflation.  
\section{Big Bang Nucleosynthesis}
While the CMB map of anisotropies can be considered as a photograph 
of the Universe when it was 380 kyr old, the observed abundances of the 
light nuclides $^1$H, $^2$D, $^3$He, $^4$He, and $^7$Li represent the most 
ancient archaelogical document about the history of the Universe
\cite{BBN}. These light
elements were cooked in nuclear reactions when the Universe was about 3
minutes old. The remaining elements is the work of the stars with a 
little help of cosmic rays spallation. The first to propose a Big Bang
nucleosynthesis was G. Gamow in 1946. In fact, a glance at the helium 
abundance: 24\% in weight against 76\% for hydrogen, tells us that so much 
helium can not have been produced by stars. For example, assuming that
the age of the Milky Way is $10^{10}$ yr and that it has been radiating
all the time at its current power $L_{\mathrm{MW}} = 4 \times 10^{36}$ W,
with all this power coming from the burning of hydrogen into helium,
this will account only for less than 1\% helium abundance.

The very early Universe is a too hostile environment for nuclei. 
When the temperature stays above a few MeV -the typical nucleon
binding energy- the photons will immediately destroy any existing nuclei.
So, nucleosynthesis has to wait until the Universe has cooled down 
enough. How much is enough? The first step for nucleosynthesis is the 
formation of deuterium through the reaction $n + p \rightarrow d + \gamma$,
and the binding energy of deuterium is 2.22 MeV. However until the
temperature does not reach below $0.1$ MeV, formation of deuterium is not 
possible due to the high specific entropy of the Universe. Put in 
other words: since the photons outnumber the baryons by a factor $10^9$, 
even well below the binding energy of deuterium, there are enough hard
photons in the high energy tail of the Planck distribution to destroy
the deuterons as fast as they are produced. Once the deuterons are able 
to survive they almost instantaneously transform into helium through the 
reaction $d + d \rightarrow ^4$He $ + \gamma$ and through other fusion 
reactions involving $^3$H and $^3$He as intermediate steps. Production
of helium is very much favoured by its comparatively high binding energy
$28.3$ MeV. So, from energetic considerations only, it could have happened
earlier, but it has to wait until deuterium is formed. This effect is called 
the \emph{deuterium bottleneck}. Finally some $^7$Li seven is also formed
in collisions of $^4$He with $^3$He and $^3$H nuclei. 
Why not higher nuclei? The reason is that there are not stable nuclei
with $A =5$ and $A=8$, and only minute quantities of the nuclei with
$A=2, 3$ are formed in the synthesis of helium. In addition, 
the Universe is cooling down very fast, and Coulomb barriers which are 
higher for higher nuclei suppress nuclear reactions. Therefore
BBN stops at this point.  
So, how do the stars manage to form the rest of the elements like $C$, $N$, 
$O$, which we the observers are made of? The answer was given by F. Hoyle.
There exists a metastable resonance of two $^4$He nuclei.
Then, if a third $^4$He nucleus meets the resonance, a $^{12}$C nucleus
is formed by a two steps chain of two particle collisions. This is 
possible if the temperature and the density are both very high, but in
the early Universe, the density and temperature continually drop, and
by the time helium has been synthesized, it is too late 
for this way of producing carbon. However in the interior of stars 
temperature and density steadily rise as the star evolves, and eventually, 
the physical conditions for the transformation of helium into carbon
-and then into heavier elements- are attained.

The nuclear and elementary particle physics needed to study BBN is well
known, and the temporal dependence of the density and temperature can be 
derived from Friedmann equations. Therefore, the light elements cosmic
abundances can be theoretically calculated and compared to the observed
ones. The synthesis of light elements is sensitive to physical conditions
for temperatures $T \lesssim 1$ MeV, corresponding to an age $t \gtrsim 1$s.
Above this temperature neutrons and protons are in thermal equilibrium
through weak reactions like  $e^{-}+p \leftrightarrow \nu_e + n$ and 
$\bar{\nu_e} + p \leftrightarrow e^{+} + n$, 
and also similar ones for the other neutrino and lepton families.
Thus the neutron abundance of protons and neutrons is fixed by the
Boltzmann factor $e^{-Q/T}$, where $Q = 1.293$ MeV is the neutron-proton
mass difference. Thus, as long as thermal equilibrium is mantained, 
the Universe is running out of neutrons very fast as it cools down. 
Indeed, if thermal equilibrium held until the deuterium bottleneck 
is surpassed, very few neutrons would survive. However this is not so 
because before that, weak interactions ``freeze-out" and neutrinos
go out of thermal equilibrium.
This happens because weak interaction cross sections scale with temperature 
as $T^2$ and non-relativistic particle densities as $T^3$, while Hubble
parameter $H$ scales as $T^2$. Therefore the weak reaction rates 
$\Gamma_w$ over the expansion rate $H$ scale as $\Gamma_w/H \propto T^3$.
Thus, eventually at some temperature, which detailed calculations show to be
$T \simeq 1$ MeV, the neutron-proton interconversion go out of thermal
equilibrium. At this point, the neutron to proton ratio is about 1/6.
From this point on, occasional weak interactions with the tails 
of the lepton and nucleon Fermi distributions, and neutron beta decay
still lower (although much more slowly) the neutron to proton ratio.
When the deuterium bottleneck is surpassed at $T = 0.1$ MeV, the Universe
is about 2 min old, and the neutron to proton ratio has fallen down
to $\simeq 1/7$. Therefore, since practically all deuterium transforms  
into helium, the helium abundance today should be about 25\% in weight.  

The observed cosmological abundance for primordial $^4$He are in the
range 23-24\%, in good agreement with the theoretical calculation
based in BBN. The observed primordial abundances for deuterium
and $^7$Li can de estimated to be in the ranges $D/H = 1-7 \times 10^{-5}$,
and $^7$Li$/H = 0.59-4.1 \times 10^{-10}$, while for $^3$He there is not 
a good estimation, which renders  $^3$He unsuitable as a cosmological 
probe. In BBN, the light elements abundances depend on the balance
between the expansion rate $H$ of the Universe during nucleosynthesis,
and the nuclear reactions rates, which in turn depend on the baryon
density. All light elements abundances can be explained with a baryon 
density given by a baryon to photon ratio $\eta_{10}$ in the range
3.4-6.9 (95\% CL). This value agrees remarkably well with the value for
the same parameter obtained from CBM acoustic peaks, and provides 
another key confirmation of the BBN theory and of the SCM.

Since the reactions building $^4$He are so rapid, $^4$He primordial
abundance depends mainly on the neutron availability when the required
temperature to surpass the deuterium bottleneck is attained, and it is
rather insensitive to baryon density. In turn, neutron availability
at deuterium bottleneck depends on how long it takes to reach this
point, i.\ e.\ on the expansion rate of the Universe. Therefore,
primordial $^4$He act as a \emph{chronometer}.  
On the other hand, the other light elements relic abundances depend 
mainly on the nucleon density and act as a \emph{baryometer}. 
In particular, primordial deuterium abundance is best known and 
depends sensitively on $\eta$. Therefore, deuterium abundance is the
baryometer of choice. 

The impressive agreement in the determination of the baryon content 
of the Universe by means of two totally independent sources of 
information: the acoustic peaks in the angular spectrum of CMB 
anisotropies, and BBN, is, beyond all doubts, a sign of the maturity 
of the SCM. However, a very important question remains unanswered:
Why the photon to baryon ratio is about $10^9$, or why
are there any baryons at all? 
We briefly adress this question in the following subsection.

\subsection{Baryogenesis}
   
Since there is a nucleon for approximately each $10^9$ photons, 
but almost no antinucleons, the observable Universe seems to have a net
baryonic number (number of baryons minus the number of antibaryons).
In addition, when the temperature of the Universe falls below 1 MeV, 
electrons and positrons annhilate into photons but not wholly. 
A small fraction of electrons in excess survive to exactly balance
the charge of the protons. Thus, there is a matter-antimatter 
asymmetry in the Universe, which is very tiny but very important for us
(otherwise we would not exist).
The exact origin of this asymmetry is still unknown, A. Sakharov
formulated three necessary conditions that must be fulfilled in
order to generate the matter-antimatter asymmetry. i) There must exist
\emph{CP violating processes}, that distinguish particle from antiparticle
interactions. Such kind of processes are known to exist in the EPSM, due to
the mixings between the three families of quarks and leptons, and
have been observed in the neutral kaons system. ii) There should
exist \emph{baryon number violating processes} that generate a net baryonic 
number. Such processes exist in Grand Unifed Theories, and also as a 
non-perturbative effect in the minimal standard model. iii) There 
should exist \emph{deviations from thermal equilibrium}. For if 
thermal equilibrium was always mantained, since particles and antiparticles
have the same mass, their abundances would be always exactly the same.
Much work has been done over the last two decades in building models 
that meet the Sakharov criteria, and predict the right amount of baryons 
in the Universe (the value of $\eta$). However, there is 
still not any conclusive explanation of baryogenesis at present
\cite{baryogenesis}.

\section{Inflation}

Despite the success of classical standard Cosmology in
explaining the expansion of the Universe, the abundances of
light elements, and the existence of a highly isotropic
cosmic microwave background; this theoretical framework 
exhibits important limitations which we will discuss in this
section.

On one hand, the assumed initial conditions for the evolution of the
homogeneous FRW background are  problematic. Thus,  observations 
favor a universe with flat or almost flat spatial sections. However, 
such an universe is an unstable solution of the cosmological
evolution equations. This implies that only a very small set of 
initial conditions could evolve into the presently observable 
Universe.  This is usually referred to as the flatness problem.
More important is the so-called horizon problem. Since the Universe
had an origin in time, the maximum distance that light can 
travel from the Big Bang until a given time (particle horizon) is
 finite. This is also the maximum size that 
a causally connected region can have at that time. 
However, when comparing the size of the particle horizon at 
matter-radiation decoupling
with the physical size of the presently observable Universe at that
time, we find that the latter was much larger than the horizon size. 
This implies that not all of our observable Universe was inside a
single causally connected region, and therefore there is no reason
to expect that background radiation photons coming from
different regions in the sky were at the same temperature. 
However, observations confirm that this is certainly the case, 
since the  temperature anisotropies are extremely small.
It is important to emphasize that these two problems arise
because we are assuming that the evolution of the Universe is
the standard one all the way down to the initial singularity. However, 
General Relativity is a classical field theory which  is expected to 
break down at very short distances where quantum effects 
would dominate.

On the other hand, as explained above, 
the formation of large scale 
structures such as galaxies
or galaxy clusters, is understood within the SCM  
as the amplification of initially small 
density perturbations, due to Jeans instability. 
However, although it is 
possible to determine the evolution of such perturbations within
classical Cosmology, the model does not provide a mechanism for the 
generation
of the primordial seeds, which are considered as an additional input.
Indeed, there is a general argument which suggests that fluctuations 
generated within
the horizon size at an early epoch by some mechanism 
(thermal fluctuations, ...) cannot be responsible for
the observed  structure at all scales. The argument reads as follows:
 consider a  
process which took place before nucleosynthesis when $T\gsim 100$ MeV. 
The mass within the 
horizon at that time was around $m_H\lsim M_\odot$. Consider also large
density fluctuations of order $\delta_H=(\delta \rho/\rho)\vert_H\sim 1$ on those scales. As larger and 
larger scales $M$ enter the horizon, the dispersion of the mass 
fluctuations 
will decrease as $N^{-1/2}$ where $N=M/m_H$ is the number of small regions
contained in the large one. Thus,  the typical size of fluctuations
on horizon scales at the time when a galactic scale with 
$M_{gal}\sim 10^{12} M_\odot$
entered the horizon,   would be 
$\delta_{gal}\sim \delta_H N^{-1/2}\lsim 10^{-6}$, which is
too small to explain the present galactic density contrast.  
Thus, the existence of structures  and temperature anisotropies at 
large scales is difficult to explain by  causal phenomena 
in classical Cosmology and suggests the presence of perturbations on 
{\it super-horizon} scales, generated by some {\it exotic} mechanism
(such as inflation). 

In fact, if we insist on solving these problems ignoring 
possible quantum gravitational effects near the Big Bang
singularity, then 
inflation can do the job \cite{Liddle}.
Although its original motivation was to 
get rid of the overproduction of  
supermassive relics (monopoles)  which were predicted by  
certain models of the very early Universe, 
it was soon realized that inflation also
provided a natural solution for the flatness and horizon
problems.

Inflation is a short phase of accelerated expansion 
in the very early Universe. If inflation lasts for a 
sufficiently long period, i.e. for a large enough 
number of e-folds:  
$N_e=\ln (a_f/a_i)$,
where $a_{f(i)}$ denotes the scale factor at the end (beginning) 
of inflation, ($N_e\gsim 50-70$ depending on the model); 
then it can be shown that the previously mentioned
problems are automatically  solved. Thus, the typical exponential growth
of the scale factor during inflation makes the spatial curvature of the Universe
to decline dramatically. 
In addition, during inflation, the physical
size of the particle horizon grows at a similar rate as the physical
distances, and as a consequence the presently visible 
patch of the Universe was at
all times well inside the causally connected region. 

Apart from the debatable importance of inflation as a solution
for the flatness and horizon problems, its major success was the 
prediction
of a (nearly) scale invariant spectrum  
of  density fluctuations on super-horizon scales, 
in agreement with observations. 
In fact, during inflation, quantum fluctuations
with sub-horizon  physical wavelengths
($\lambda\ll H_I^{-1}\sim (10^{13}$ GeV)$^{-1}$ in typical models)  
can be stretched by
the Universe expansion up to scales comparable to the size of galaxies,  
galaxy clusters (kpc-Mpc) or even larger,  at the present epoch. 
The amplitude
of the quantum fluctuations of any {\it light}
scalar field 
at horizon crossing is determined by the 
Hubble parameter $H_I$. 
Then, since this parameter is typically almost constant during inflation,
a generic prediction of any inflationary model, is the mentioned
flat spectrum of  perturbations. Such form for the
spectrum 
had been postulated many years before by Harrison and Zeldovich
 in order to explain galaxy formation.  
In addition to 
acting as seeds for structure formation as explained
above, these perturbations are also responsible for
the generation of anisotropies in the background radiation through the
already mentioned Sachs-Wolfe effect.

 It can be seen that those large scales 
which became
larger than the Hubble radius (exit the horizon) at the beginning
of inflation, re-entered the Hubble radius later, whereas the 
shorter wavelengths re-entered sooner, following 
a LOFI (last out, first in) scheme. Linear perturbation
theory shows that once a given scale has entered
the horizon, the corresponding density fluctuation can grow 
linearly with the scale factor $a$ in the matter dominated era,
whereas the growth is only logarithmic in the radiation dominated one.
Thus, the scale $\lambda_{eq}\simeq 13 (\Omega_M h^2)^{-1}$ Mpc
 corresponding to the size of the Hubble horizon at matter-radiation 
equality separates out the two behaviours.
  Therefore, the perturbations with shorter wavelengths have had
more time to grow since reentering than the larger ones. 
So, there are  
definite predictions of inflation for structure formation
and CMB anisotropies. 
Since the primordial spectrum is flat, i.e. perturbations
have the same amplitude at all scales, the
supression factor (ignoring possible
non-linear effects) in the density constrast will be given by 
$(a_{eq}/a)=(\lambda_{eq}/\lambda)^2$ 
for $\lambda>\lambda_{eq}$, whereas it will be only logarithmic
for shorter wavelengths. The predictions agree reasonably well with
observations (see Fig. 1). In addition, the presence of the  so 
called Sachs-Wolfe plateau 
in the large angular scales region of the CMB  
power spectrum (see Fig. 3), corresponding to wavelengths
of thousands of Mpc, can  be traced back also to the 
inflationary prediction.

Although inflation is at present the only viable scenario 
of the early Universe, unfortunately it is not a
complete theory. The mechanism responsible 
for the accelerated expansion is unknown. Most of the inflationary
models are based on the existence of a hypothetical scalar field 
called {\it inflaton} (either fundamental or effective) whose potential energy 
density dominates at early times,  
acting as an effective cosmological constant. However, only extensions of
 the Standard Model of elementary particles, such as supersymmetry, supergravity
or string theory could naturally accomodate a scalar field with
the required properties. These models are not free from 
difficulties either, since the required smallness of the 
slow-roll parameters
for the inflaton potential,
which is needed to fit observations, can only 
be maintained in particular cases.

The increasing observational precision in CMB anisotropies and
 large scale structure has allowed
to improve the constraints on inflationary models. In particular,
the primordial curvature power spectrum predicted by inflation can
be parametrized as:
\begin{eqnarray}
P_{R}(k)=A_S^2\left(\frac{k}{k_0}\right)^{n_s-1}
\end{eqnarray}
where $A_S$ is the amplitude of scalar metric perturbations, 
$n_s$ is the spectral index and $k_0$ is the normalization
scale. Observations seem to be compatible
with a simple power-law, gaussian, adiabatic spectrum, although a 
small contribution from
isocurvature modes cannot be excluded. The combined analysis of CMB data
from WMAP satellite, ground-based detectors such as CBI and ACBAR, 
the 2dF galaxy redshift survey and Ly-$\alpha$ forest information,
 provides the following  68$\%$ C.L. results 
at the $k_0=0.05$ Mpc$^{-1}$ scale:
\begin{eqnarray}
A_S&=&(4.1\, -\, 5.0 )\cdot 10^{-5}\nonumber \\
n_s&=&0.90\, -\, 0.96 \nonumber \\
dn_s/d\ln k&=&-(0.049\, -\, 0.015)
\end{eqnarray}

We see the agreement with the Harrison-Zeldovich prediction 
$n_s\simeq 1$, and the small running of $n_s$ with the scale.
Although at present, these results do not exclude any kind
of inflationary model, some particular form of the inflaton
potential could be disfavored in the near future when new
data are available.

Apart from scalar perturbations, inflation also predicts the 
generation of a gravity wave background,  characterized too by its
power spectrum:
\begin{eqnarray}
P_T(k)=A_T^2\left(\frac{k}{k_0}\right)^{n_t}
\end{eqnarray}
The amplitude of tensors is usually compared with the scalar
amplitude in the tensor-scalar ratio $r=A_T^2/A_S^2$. 
In single-field models of inflation, the tensor spectral index
is related to $r$ thorugh the consistency condition $n_t=-r/8$, so that
the number of independent parameters can be reduced to 
$(A_s,n_s, r, dn_s/d\ln k)$. Since no gravity
wave mode has been detected yet, the previous combined
analysis  gives only the constraint: $r<0.9$ at the 95 $\%$ C.L.
(a very recent fit from SDSS+WMAP+Ly-$\alpha$+SNIa data has obtained
a better bound, $r<0.45$ at the 95 $\%$ C.L. \cite{SDSSnew})
Things can improve if we take into account the effect of 
CMB polarization. Gravity waves produce {\it magnetic} components
of polarization (B-modes) which are not
produced by scalar perturbations. Planck satellite, 
with polarized detectors,  is expected to measure $r$ with error bars 
around $0.13$. 

The possibility of testing different inflationary models 
by future experiments will
open a fascinating window to the physics of the  early 
Universe. Indeed, 
quantum
fluctuations generated during inflation 
well inside the Hubble radius have  typical
wavelengths much smaller than those probed by current particle
 accelerators. In other words, inflation tells us that the 
CMB temperature
anisotropies and the large scale structures that we observe today,  
are the signals of very high-energy physics in the sky.

\vspace{.5cm}

\centerline{ 
\rule[.1in]{6cm}{.002in}}


\begin{table}[h]
\begin{center}
\centerline{\large{\bf The Cosmic Inventory}}
\vspace{.2cm}
\renewcommand{\arraystretch}{1.5}
\begin{tabular}{|c|c|}
\hline\hline   
Hubble parameter  & 
$h=0.71^{\;+0.04}_{\;-0.03}$ \\
Baryon density  & $\Omega_B=0.044\pm 0.004$\\ 
Matter density  & $\Omega_M=0.27\pm 0.04$\\  
Dark energy density  & $\Omega_\Lambda =0.73\pm 0.04$\\ 
Total energy density  & $\Omega_{tot}=1.02\pm 0.02$\\ 
Neutrino density  & $\Omega_\nu h^2\leq 0.0076$ (95 $\%$ C.L.)\\ 
Dark energy equation of state & $ w< -0.78$ (95 $\%$ C.L.)\\
\hline 
\hline  
\end{tabular}
\vspace{.2cm}
\caption{\footnotesize{Cosmological parameters with 
$68 \%$ C.L. intervals (except as otherwise stated). 
Results from the combined fit of WMAP, CBI, ACBAR, 2dF and 
Ly-$\alpha$}}
\end{center}
\end{table} 
 {\bf Acknowledgements:}  This
work has been partially supported by the DGICYT (Spain) under the
project numbers FPA2000-0956 and BFM2002-01003.

\end{document}